\documentclass[submission,copyright,creativecommons]{eptcs}
\providecommand{\event}{FESCA 2013}

\usepackage[usenames,dvipsnames]{color}
\usepackage[T1]{fontenc}
\usepackage{stmaryrd}
\usepackage{pifont}
\usepackage{alltt}
\usepackage{amssymb}
\usepackage{amsmath}
\usepackage{txfonts}
\usepackage{float}
\usepackage{multirow}
\usepackage[pdftex]{graphicx}
\usepackage{verbatim}
\usepackage{colortbl}
\usepackage{xspace}
\usepackage{tikz}
\usepackage{xypic}

\newcommand{\highlight}[1]{\colorbox[gray]{0.9}{#1}}

\newcommand{\polylarva}{\newlarva}
\newcommand{\newlarva}{poly\textsc{Larva}\xspace}
\newcommand\OpenEmm{\texttt{OpenEmm}\xspace}
\newcommand\mboxit[1]{\mbox{\textit{#1}}}

\newcommand{\ie}{\textsl{i.e.,}\xspace}
\newcommand{\eg}{\textsl{e.g.,}\xspace}

\newcommand{\etc}{\textsl{etc.}\xspace}

\floatstyle{ruled}
\newfloat{program}{thp}{lop}[section]
\floatname{program}{Program}

\newenvironment{newcode}{\footnotesize\begin{alltt}}{\end{alltt}\normalsize}

\newtheorem{example}{Example}[section]

\def\authorrunning{C. Colombo, A. Francalanza, R. Mizzi, and G. J. Pace}

\def\titlerunning{Extensible Technology-Agnostic Runtime Verification}

  \title{Extensible Technology Agnostic Runtime Verification} 
  \author{Christian Colombo\qquad Adrian Francalanza \qquad Ruth Mizzi \qquad Gordon J. Pace
  \institute{Department of Computer Science\\ University of Malta}
  \email{\{christian.colombo|adrian.francalanza|rmiz0015|gordon.pace\}@um.edu.mt} }
 
\begin{document}
\maketitle

\begin{abstract}
With numerous specialised technologies available to industry, it has become increasingly frequent for computer systems to be composed of heterogeneous components built over, and using, different technologies and languages. While this enables developers to use the appropriate technologies for specific contexts, it becomes more challenging to ensure the correctness of the overall system. In this paper we propose a framework to enable extensible technology agnostic runtime verification and we present an extension of \newlarva, a runtime-verification tool able to handle the monitoring of heterogeneous-component systems. The approach is then applied to a case study of a component-based artefact using different technologies, namely C and Java. 
\end{abstract}

\section{Introduction}


Component-based approaches to software and service design are becoming more widespread, allowing for heterogeneous systems to be compartmentalised into components; these components encapsulate their internal behaviour while revealing interfaces through which other components may interact.  This approach to system organisation has facilitated the construction of large complex systems, where each component is allowed to internally employ 
different technologies, from operating systems and hardware, to programming languages.

However, the sheer complexity of the systems constructed, together with the decentralised nature of how these heterogeneous systems are developed,  creates new potential points of failure.  This, in turn, increases the need for some form of correctness verification. Runtime monitoring \cite{rv} has been shown to be a viable solution to the verification of large complex systems --- by limiting the analysis to the actual current runtime path, the approach is tractable (for a reasonable choice of specification logic), while still guaranteeing detection (albeit at runtime) of property violations. Especially in the context of open systems, where correctness is also partly dependent on interaction with the environment, this approach has proved to be viable and scalable to real-life industrial systems \cite{rvIndustry2,rvIndustry}.




An important research question in the field of runtime verification has been how to extend the approach to handle non-monolithic systems. Much of the work so far has been limited to a high-level view of such systems, treating components as \emph{black boxes} and focusing instead on the verification of the component interactions and on strategies for engineering monitoring in such distributed settings. To date, there has been little work that attempts to push verification \emph{inside} the components so as to verify their inner workings as part of the wider system.   This poses additional challenges to runtime verification, both practical and theoretical, such as the need for a standardised framework for generating events to be monitored (irrespective of the underlying technology used by the monitored components), or the adaptation of the monitoring logic (specifying correctness properties) to distinct underlying technologies used by different components, which may invariably lead to different semantic interpretation of said logics. 

A typical example would be an online betting system, which is inherently component-based (\eg the web-portal subsystem, the billing subsystem, the fraud detection subsystem \etc), and where each component may use different technologies and resources.  Typical operations in this system, such as an online-betting transaction, may go through different components ranging from the company's transaction database, to an internal certified logging component kept for legal purposes, to an external bank system.   Whereas existing verification technologies would typically only be able to monitor external units as black boxes, thus only referring to the interactions taking place, consistency and correctness properties of such betting transactions may depend on the inner workings of internal components. For instance, one may want to ensure that the value written in the certified logging component matches that written in the company's transaction database. Such component-spanning properties have to be instrumented on different components, necessitating direct interaction with the underlying technologies of each component. Despite this, the present lack of tools supporting such technology-heterogeneity  is a major stumbling block towards a full adoption of runtime verification techniques for component-based systems.



In this paper we present a novel runtime verification framework which supports the monitoring of component-based systems, possibly using different technologies. The approach has been implemented as an extension to the runtime verification tool \polylarva \cite{conf/sefm/ColomboFMP12}, to support the generation and instrumentation of separate monitoring code for multiple components in a system, and from a single property specification. Furthermore, the property specification language has been designed to be technology and programming-language agnostic, and hence allows reusability across different technologies.  To evaluate the proposed component-based monitoring framework, \polylarva\ has been applied to \OpenEmm, an open source web-based e-mail marketing tool.

In the rest of the paper, we first discuss the various design options for the runtime verification of component-based systems; see Section \ref{sec:des}. In Section \ref{sec:arch}, these issues are then discussed in the context of multi-technology component-based systems where the monitor needs to access the components' internal states. Subsequently, Sections \ref{sec:spec} and \ref{sec:plugs} describe how we support our design decisions through an extensible monitoring framework, \newlarva. In Section \ref{sec:cs} we apply the approach on a case-study, after which we conclude and discuss related work. 


\section{Challenges of Component-Based Monitoring}\label{sec:des}

From a monitoring perspective, component-based systems pose particular challenges which go beyond those present in monolithic systems. While the well-behaved individual components can be dealt with locally, properties and specifications which span across components so as to express the correctness of the system as a whole raise various issues, both pragmatic and conceptual.  In particular, the cross-cutting nature of monitoring, which may require access to the internal state of different components, raises issues regarding the architectural design of the monitors, particularly, if components are built using different technologies.

\noindent Two important issues, which influence choices in the monitoring architecture are:

\begin{description}
\item[Orchestrated vs. choreographed monitoring:] 
Different approaches have been proposed in the literature as regards to the locality (physical or conceptual) of the monitors with respect to the different components in a system. In an orchestrated monitoring approach (\eg see \cite{Gan:2007:RMW:1321211.1321217,1615040,springerlink:10.1007/978-3-540-77395-56}), the monitor is a separate unit from the rest of the system, but with privileges to enable it to listen to the behaviour of the other parts of the system and pass judgement about them. On the other hand, a choreographed approach (\eg see \cite{choreo1,choreo2,choreo3}) extends the different parts of the system such that each part \emph{locally} eavesdrops on its own behaviour, ensuring correct behaviour, and communicating with other monitors (belonging to other parts of the system) whenever synchronisation or information regarding the other parts is required. Approaches to split monitors in this manner using a static \cite{choreo1} or dynamic \cite{choreo0} orchestration have been proposed. Although most of this work focuses on distributed systems, the classification applies equally well to general component-based systems. Orchestration is usually easier to instrument and setup, but adds dependencies between components which may not be desirable. On the other hand, a choreographed approach respects locality, but at the cost of more complex instrumentation of monitoring code and specification slicing. Especially in the case of heterogeneous component-based systems with components using different technologies, local instrumentation can prove to be particularly challenging, since the monitoring tool has to be able to instrument code written in different languages.

\item[Intrusiveness of monitoring:] 
Another independent choice is the level of abstraction at which the monitors can eavesdrop the behaviour of the system. Much work, especially in setting of services, focuses on specifications of messages passed between components \cite{Gan:2007:RMW:1321211.1321217,1615040}. This black-box approach ensures that the instrumentation of monitors is relatively straightforward since they only need to hook to the communication channels and process the behaviour appropriately. However, the approach has serious limitations when specifications refer to the components' local states or internal events since, without re-engineering to expose such states and events, their monitoring would not be possible. In a component-based system setting, such properties would, for instance, be required to ensure data-consistency across components. A more intrusive approach, to enable direct access to the state of the system components poses challenges, especially when the components are built using different technologies.
\end{description}

Note that these two issues are largely independent of each other. For instance both black-box orchestrated monitoring \cite{Gan:2007:RMW:1321211.1321217,1615040} and intrusive orchestrated monitoring \cite{springerlink:10.1007/978-3-540-77395-56} approaches have been proposed in the literature. In the next section we investigate how challenges in extending the existing approaches to technology-diverse component-based systems can be addressed.

\section{Intrusive Monitoring of Heterogeneous Component-Based Systems}
\label{sec:arch}

For a number of applications, checking for specification violations  through eavesdropping on component communication suffices.   However, there are situations (\eg see Section \ref{sec:cs}) where more intrusive monitoring is required. In the case of systems built from components using different technologies, intensional form of monitoring poses the following challenges. 

\subsection{Monitoring Architecture}

Supporting technology agnosticism using an intrusive choreographed approach limits \emph{extensibility} since each technology would have to be potentially coupled with another one. In contrast with an orchestrated approach, adding support for a new technology only requires adding means to instrument code which handles communication with the central monitor. 

For this reason, we adopt a centralised monitor which receives events and processes them, thus making it a largely orchestrated approach --- with the verification unit of the tool being independent of the the other components. For every supported technology, the tool can intrusively instrument event generation into the components.  

\begin{figure}
\centerline{\scalebox{0.65}{\input{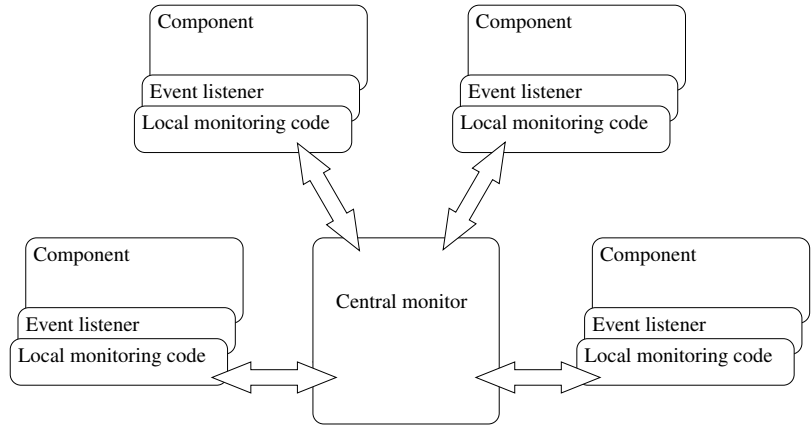}}}
\caption{Architecture of instrumented monitors on components}
\label{f:architecture}
\end{figure}

Since a purely orchestrated, yet intrusive, approach may result in breaches of data and control encapsulation within components, our framework gives full control to the user as to the extent of intrusiveness of the monitoring logic onto the system-side \cite{conf/sefm/ColomboFMP12}. This allows for manually programmed choreography, without sacrificing extensibility, since inter-component coordination still happens through the central orchestrated monitor. Fig.\ \ref{f:architecture} shows the general architecture of the system after the monitoring parts are instrumented onto the components.

\subsection{Extensible Technology Support}

Through the general monitoring architecture of Fig.~\ref{f:architecture}, adding support for new technologies requires the possibility of instrumenting monitors into components built using that technology. The responsibility of the \emph{local} monitoring code is to $(i)$ generate events to send to the central monitor; and $(ii)$ execute any local monitoring code specified in the properties. 

Since it is desirable to support cross-component properties, it is crucial to support monitoring instrumentation from a single specification script; this is the approach adopted for \polylarva. Furthermore, since extensibility to further technologies is also an essential feature, we have separated the monitoring instrumentation into different parts $(i)$ the central verification code is generated from the global parts of the specification using a language-independent part of the runtime verification tool; and $(ii)$ for each different technology, a separate tool is provided, which generates the automated instrumentation scripts from a subpart of the specification (as tagged by the user).

\begin{figure}
\centerline{\scalebox{0.5}{\input{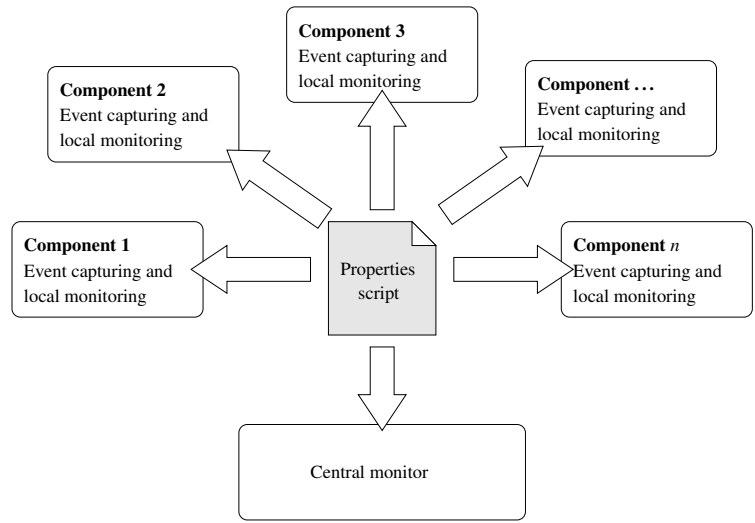}}}
\caption{Splitting the specification script across components}
\label{f:specification-components}
\end{figure}

The workflow for the usage of \newlarva is depicted in Fig.\ \ref{f:specification-components}: $(i)$ the specification script is passed through the language agnostic part of the tool to produce the central monitor (bottom arrow of the figure); and $(ii)$ for each different component, the appropriate technology tool of \newlarva processes the specification to produce the instrumentation instructions for event extraction and local monitoring on that particular component (top arrows in the figure). 

The actual instrumentation typically takes place by processing the scripts produced by the \newlarva language-dependent compilers, and subsequently using additional external tools (such as aspect-oriented programming compilers).

\subsection{Replicated Monitors and Language Agnosticism}

Since most systems describe multiple instances of an abstract concept (\eg multiple users, accounts, sessions, \etc) frequently, one would desire to replicate a property for each instance. For example, for each bank transaction, one may want to set up a monitor for a property which states that the incoming and outgoing balances cancel out. In practice, the way such concepts are encoded depends on the technology being used. For instance, if the system is written in Java, a transaction may be encoded as an object, while if written in Erlang,\footnote{http://http://www.erlang.org} it may correspond to a separate actor process handling the transaction. While monitoring tools for single-technology monolithic systems associate such replication in correspondence with  the technology, in a language agnostic system one needs to be more general. In component-based systems, this poses further challenges when the concept's lifetime spans across different components. To support such replication, one solution is to demarcate concept instances' lifetimes by identifying events marking their start and those marking their end. Furthermore, all events related to such an instance are tagged with an identifier which indicates to which instance they belong.

In the bank transaction example, we would identify a call to \texttt{initialiseTransaction(transid)} to be the starting point, while \texttt{concludeTransaction(transid)} with the same parameter to be its end. Moreover, any calls to \texttt{transferFunds(transid,\ldots)} would be associated with the instance of the monitor which was started with the transaction identity passed as a parameter.

\section{\newlarva Specifications}\label{sec:spec}

At the simplest level, our monitoring framework, \newlarva, uses a  guarded-command style specification language. Properties are expressed as a list of rules of the following form:

\centerline{$\mboxit{event} \;\mid\; \mboxit{condition} \;\mapsto\; \mboxit{action}$}

Whenever an \emph{event}  (possibly having parameters) is generated by the system, the list of monitor rules is scanned  for rule matches relating to that event. If a match is found, the expression specified in the \emph{condition} of the rule is evaluated and, if satisfied, the \emph{action} is triggered\footnote{Note that the condition and action may consist of any valid code of the underlying technology being used --- Java in the case of the previous version of \polylarva, and any technology in the case of the modified \polylarva presented in this paper.}.

\begin{example}
\label{example:register_then_pay}

Consider a scenario in which one desires to check that an online payment on a web-based system is carried out after the credit card used has been registered --- tagging the customer as untrusted if this rule is violated. 
%
%
This may be expressed in terms of the rules enclosed within the \texttt{rules} block, \ding{172}, in Program~\ref{prg:regpay}. 
In these rules, \texttt{register} and \texttt{pay} are system events (method calls), parametrised by the values \texttt{customer} and \texttt{card}; 
$\lnot$\texttt{registeredCards[card]} is a condition, while 
\texttt{registeredCards[card] := true} and \texttt{setUntrusted(customer)} are actions triggered by the runtime monitor. 
Note that to monitor the property for each system user, we define the event \texttt{newSession} as the point triggering the replication of the rules (\ding{173}). Consequently, the events declared within this scope must all define a variable \texttt{customer} which binds the event with a particular monitor instance. Finally, \ding{174} marks the end of the context upon event \texttt{endSession}. 

 The \texttt{state}, \texttt{conditions} and \texttt{actions} blocks define specification-related monitor state, conditions and actions respectively that are used in the \texttt{rules} section as macros.  The \texttt{events} block highlights those specific points which, during system execution, should trigger the monitoring functionality. 


\begin{program}
\begin{newcode}
\ding{173} \highlight{upon (newSession(customer))} \{
    state \{ 
      boolean[] registeredCards;   
    \}
    events \{
        newSession(customer) = \{customer.logIn();\}
        register(customer,card) = \{customer.registerCard(card);\}
        pay(customer,card) = \{customer.makePayment(card);\}
        endSession(customer) = \{customer.logOut();\}
    \}
    conditions \{        
      isRegistered(card) = \{ registeredCards[card] \} 
    \}
    actions \{
      setUntrusted(customer) = ... 
      registerCard(card) = \{ registeredCards[card] := true \}
    \}
 \ding{172} \highlight{rules} \{
      register(customer,card) \verb+->+ registerCard(card);
      pay(customer,card) \verb+\+ !isRegistered(card) \verb+->+ setUntrusted(customer);
   \ding{174} endSession(customer) \verb+->+\highlight{Done;}
    \}
  \}
\end{newcode}
\caption{Monitoring customers attempting payments with unregistered cards}
\label{prg:regpay}
\end{program}

\end{example}

While events are a result of the execution path being followed by a system, and thus specific to the system's technology, the conditions and actions defined in a rule are typically specific to the runtime monitoring states and are thus independent from the system's technology. This means that the evaluation of rule actions and conditions need not run on the system being monitored and can safely be moved onto separate resources. However, there are exceptions to this since conditions may also query the system state, and actions may alter the system state. Unfortunately, it is not straightforward to automatically delineate system-dependent elements from purely monitoring elements. 
For this reason, our language enables the user to explicitly specify  this separation through appropriate constructs\footnote{A lengthy discussion of how one could use the distinction between system-side and monitor-side monitoring to optimise efficiency can be found in \cite{conf/sefm/ColomboFMP12}.  Note, however, that in the previous work we assume that the system- and monitor-side are of the same technology.}.

\begin{example}
Further to Example~\ref{example:register_then_pay}, consider the case where instead of keeping track of which cards have been registered or not within the monitor state, we query the system state (which keeps track of card registration anyway). In this case the condition which checks whether a credit card is registered, is performed on the system state by being tagged as \texttt{systemSide},  \ding{175}, as opposed to \texttt{monitorSide},  \ding{176}. Note that the rest of the monitoring elements are still performed on the monitor side and are thus marked as such. 

%
      
\begin{program} 
\begin{newcode}
upon (newSession(customer))  \{
    events \{
        newSession(customer) = \{customer.logIn();\}
        register(customer,card) = \{customer.registerCard(card);\}
        pay(customer,card) = \{customer.makePayment(card);\}
        endSession(customer) = \{customer.logOut();\}
    \}
    conditions \{
    \ding{175} \highlight{systemSide} \{ isRegistered(card) = \{ registeredCards[card] \} \}      
    \}    
    actions \{
    \ding{176} \highlight{monitorSide} \{ setUntrusted(customer) = ... \}
    \}
    rules \{
      pay(customer,card) \verb+\+ !isRegistered(card) \verb+->+ setUntrusted(customer);          
      endSession(customer) \verb+->+ Done;
    \}
  \}
\end{newcode}
\caption{Monitoring activity between two components}
\label{prg:twosyswithsysSide}
\end{program}

\end{example}


The tagging of monitor side and system side evaluation of monitoring logic suffices for a monolithic system. However, when a system is composed of heterogeneous components residing on different technologies, tagging has to be more comprehensive and distinguish between components. There are two aspects to this: $(i)$ states, conditions, and actions may reside within different components; and $(ii)$ the events may now also arise from different components. 
These features were not previously supported in \polylarva but are now supported in the extended version presented in this paper. 
The following example, demonstrates the use of extended tagging in the context of multiple components. 

\begin{figure}[t]
\centerline{\scalebox{0.65}{\input{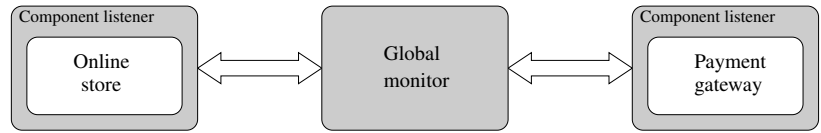}}}
\caption{Centralised monitoring for online payments}
\label{fig:distsystem}
\end{figure}

\begin{example}
As an example, we note that many online stores normally incorporate functionality offered by payment gateway web services to validate credit card details and accept transactions. In such cases, we may want to ensure that the payment details input by the customer on the online system are the same details received by the payment gateway. A possible monitoring setup for such a scenario is as illustrated in Fig.\ \ref{fig:distsystem} where a system side monitor associated with the online store system, notifies the runtime monitor about a payment transaction. Program~\ref{prg:twosys} highlights how the proposed setup is facilitated through the use of the constructs supplied by the \newlarva\ specification language. The payment transaction event \ding{177} is identified as being an event that will occur on one particular system component through the use of a user-defined label that identifies the component as \texttt{store}. The event is also parametrised with the credit card details entered to the system, thus ensuring that the global monitoring component can maintain a copy of these values. Upon receipt of an authorisation request, the system side monitor associated with the payment gateway service, will communicate with the global monitoring component in order to trigger validation of the card details. This communication is identified through the definition of the event \texttt{receiveDetails} \ding{178} which specifies that its source is one labelled as \texttt{paymentService}, and therefore a different process from that which triggered other events. 

\begin{program}
\begin{newcode}
upon (newSession(customer))  \{
    state \{ 
      monitorSide \{int cardNo;  \} 
    \}
    events \{
        \ding{177} \highlight{event@store} newSession(customer) = \{customer.logIn();\}
        \ding{177} \highlight{event@store} register(customer,card) = \{customer.registerCard(card);\}
        \ding{177} \highlight{event@store} pay(customer,card) = \{customer.makePayment(card);\}
        \ding{178} \highlight{event@paymentService} receiveDetails(customer,card) = \{incomingPayment(customer, card);\}
        \ding{177} \highlight{event@store} endSession(customer) = \{customer.logOut();\}
    \}
    conditions \{
      monitorSide \{ validateCardDetails(card) = cardNo == card; \}
      
   \ding{177} \highlight{systemSide@store} \{ isRegistered(card) = \{ registeredCards[card] \} \}      
    \}    
    actions \{
      monitorSide \{ setUntrusted(customer) = ... \}
      monitorSide \{ saveCardDetails(card) = cardNo := card.cardNo; \}
      monitorSide \{ reportError = ...\}
    \}
    rules \{
      pay(customer,card) \verb+\+ !isRegistered(card) \verb+->+ setUntrusted(customer);          
      pay(customer,card) \verb+\+ isRegistered(card) \verb+->+ saveCardDetails(card);
      receiveDetails(customer,card) \verb+\+ !validateCardDetails(card) \verb+->+ reportError();
      endSession(customer) \verb+->+ Done;
    \}
  \}
\end{newcode}
\caption{Monitoring activity between two components}
\label{prg:twosys}
\end{program}


\end{example}



\section{Extending \newlarva for New Technologies}\label{sec:plugs}

Due to the intrusive nature of monitoring in \newlarva, technology-specific plugins have to be implemented to support monitoring of a range of systems written in different programming languages. The resultant monitoring code should support functionality that allows for the generation of events, which are sent to the central monitor, and the execution of any local monitoring code. 

\begin{description}
\item[Eliciting events] To extract monitoring information from the components, in the sample plugins implemented we have opted for aspect-oriented programming (AOP) extensions. AOP has been heavily used for implementing runtime verification tools (\eg see \cite{Colombo:2009:DER:1614485.1614501,Meredith:2012:OMR:2215862.2215866}) and mature AOP solutions are available for a large number of programming languages. Still, it is up to the user to decide the approach to elicit events from a component, depending mostly on the underlying technology. 

\item[Communication to and from global monitor] Standard TCP-based socket communication has been chosen to support the transfer of messages between the runtime monitors. The decision of socket-based communication, as opposed to techniques such as remote method invocation (RMI)\footnote{http://www.oracle.com/technetwork/java/javase/tech/index-jsp-136424.html} or .NET
remoting\footnote{http://msdn.microsoft.com/en-us/library/kwdt6w2k(v=vs.71).aspx} is centred mainly around the need of a communication solution that is \emph{technology agnostic}. In addition, sockets are low-level enough to be optimisable, and require little-to-no configuration as opposed to communication standards such as CORBA\footnote{http://www.corba.org/}.

\item[Support for component-specific monitoring functionality] Allowing the evaluation of some of the monitoring logic to take place on the component side requires that a monitoring state is maintained at runtime at the component side. The structure chosen to represent the component-specific monitor and its state is heavily dependent on the language in question. 
\end{description}

To support the specification of new language plugins, \newlarva provides a language compiler API which potential plugins have to implement. 
For example one of the API methods is the \emph{eventToAspect} method which given an event from the \newlarva script, generates the corresponding AOP code which would trigger on the specified system method call. Program \ref{prg:plugin} shows code excerpts from the C plugin implementation which generated ACC\footnote{https://sites.google.com/a/gapp.msrg.utoronto.ca/aspectc/} (AspeCt-oriented C) aspect code by appending a string builder. 

\begin{program}
\begin{newcode}
@Override
public String eventToAspect (Context c, String name, Event e) \{
   ...
//event matches before the method call
   bldr.append("before ( call (");
	
// method signature
   bldr.append(e.getTarget().getType());
   bldr.append(e.getmethodName());

// method arguments
   bldr.append("(...) )");

// bind variables to arguments
   Iterator<Variable> varIter = e.params.values().iterator();
   if (varIter.hasNext()) \{
      ...
      bldr.append(" && args (" + v.getName() + ")");
   \}
   ...
   return bldr.toString();
\}
\end{newcode}
\caption{Implementing the C \newlarva plugin}
\label{prg:plugin}
\end{program}




Overall, the effort required for the development of a new plugin is non-trivial, but it is greatly facilitated by the API that directs the programmer through all the steps required. The style of the programming language to be supported and the availability of an aspect-oriented solution for the language also play a part in determining the ease with which the plugin can be created. 
To support the \OpenEmm case study (introduced in the next section), two plugins were implemented: one for Java and another for C. In both cases, relatively mature AOP extensions (AspectJ\footnote{http://eclipse.org/aspectj/} and ACC) were available for use, thus leading to the translation from event specification to aspect code to be a straightforward exercise. 

Apart from using a different AOP technology, we note that the C plugin differed substantially from Java. 
For instance, while the Java plugin implementation adopts the approach of creating a class to represent the runtime monitor (its state as class attributes and its system-side functionality as class methods), the C implementation uses records to maintain monitor state, and generic functions to implement the system-side functionality. 
These differences mainly arise because the two languages belong to different programming paradigms. For languages of the same paradigm we expect the plugins to be more similar. Hence, the sample plugins that are available to date (Java and C) can be used as the templates for the creation of a number of other language plugins that implement the same programming paradigm.  For instance, we expect the development of language plugins for languages such as .NET and C++ to be similar to the Java plugin that is already available. On the other hand, implementing plugins for considerably different technologies such as Erlang and PHP may not prove to be  as straightforward. Nevertheless, ongoing work on the development of such plugins suggests that the effort required for their development is comparable to constructing a plugging for a particular paradigm form scratch (\eg in Erlang, the runtime monitor would probably be expressed as an actor process).

\section{Case Study}\label{sec:cs}

\newlarva has been used to monitor \OpenEmm\footnote{http://www.openemm.org/}, an open source, web-based tool for email marketing. The tool provides facilities to administer mailing lists, create email shots, schedule automatic email sending, track sent emails, and manage bounced emails. \OpenEmm claims over hundreds of thousands of downloads to date and has a user base that includes a number of prominent large-scale companies.
The tool is ideal as a case study for \newlarva's technology-agnostic setup due to its component-based setup and the hybrid technologies it adopts: the core of the system is written in Java  
while performance-sensitive components are written in C. 

\begin{figure}[t]
	\centering
  	\scalebox{.75}{
       \includegraphics{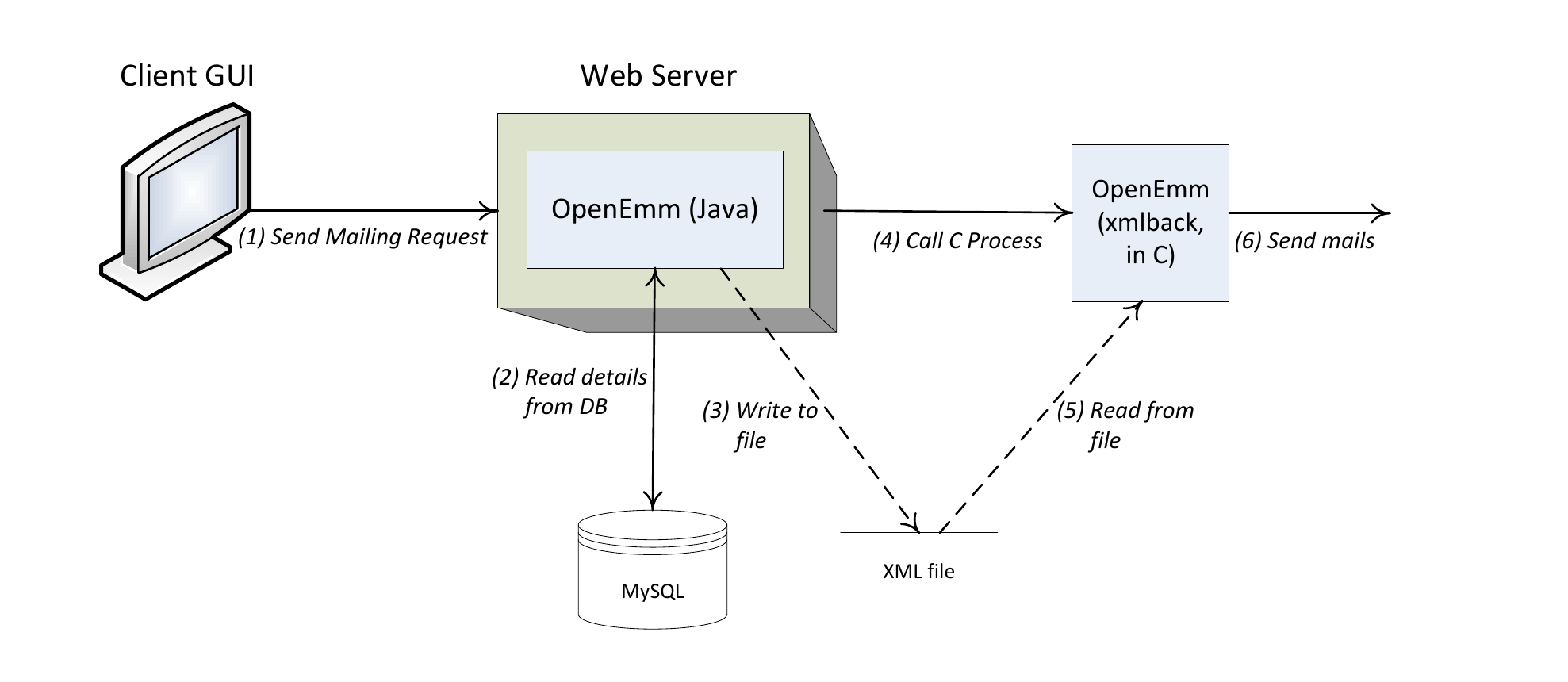}}
	\caption{Partial architecture of the \OpenEmm framework}
	\label{fig:oeArch}
\end{figure}

The case study focuses on monitoring the behaviour of \OpenEmm to send out a customised email shot. The tool provides users with a content management interface that allows easy customisation and creation of email templates. When an email is sent out to a mailing list, each individual recipient will receive a personalised email, built by an automated process that customizes each email based on the given template. 
This functionality is a result of a process which flows across two components: a Java component which retrieves the email template and mailing list information from the database and collates them in an XML file, and a C component which receives this information and produces the final emails. This control flow is displayed using sequentially numbered arrows in Fig.\ \ref{fig:oeArch}.

The setup looks out for the possibility of discrepancies between the two components: \eg the XML file getting corrupted, or the database gets updated resulting in outdated emails being sent.  This concern is addressed by (intrusively) monitoring the components' states and identifying any inconsistencies which may occur. The rest of this section elaborates on how this has been achieved.

\subsection{Specifying Properties for \OpenEmm}

\begin{figure}[t]
	\centering
		\scalebox{.75}{		    \includegraphics{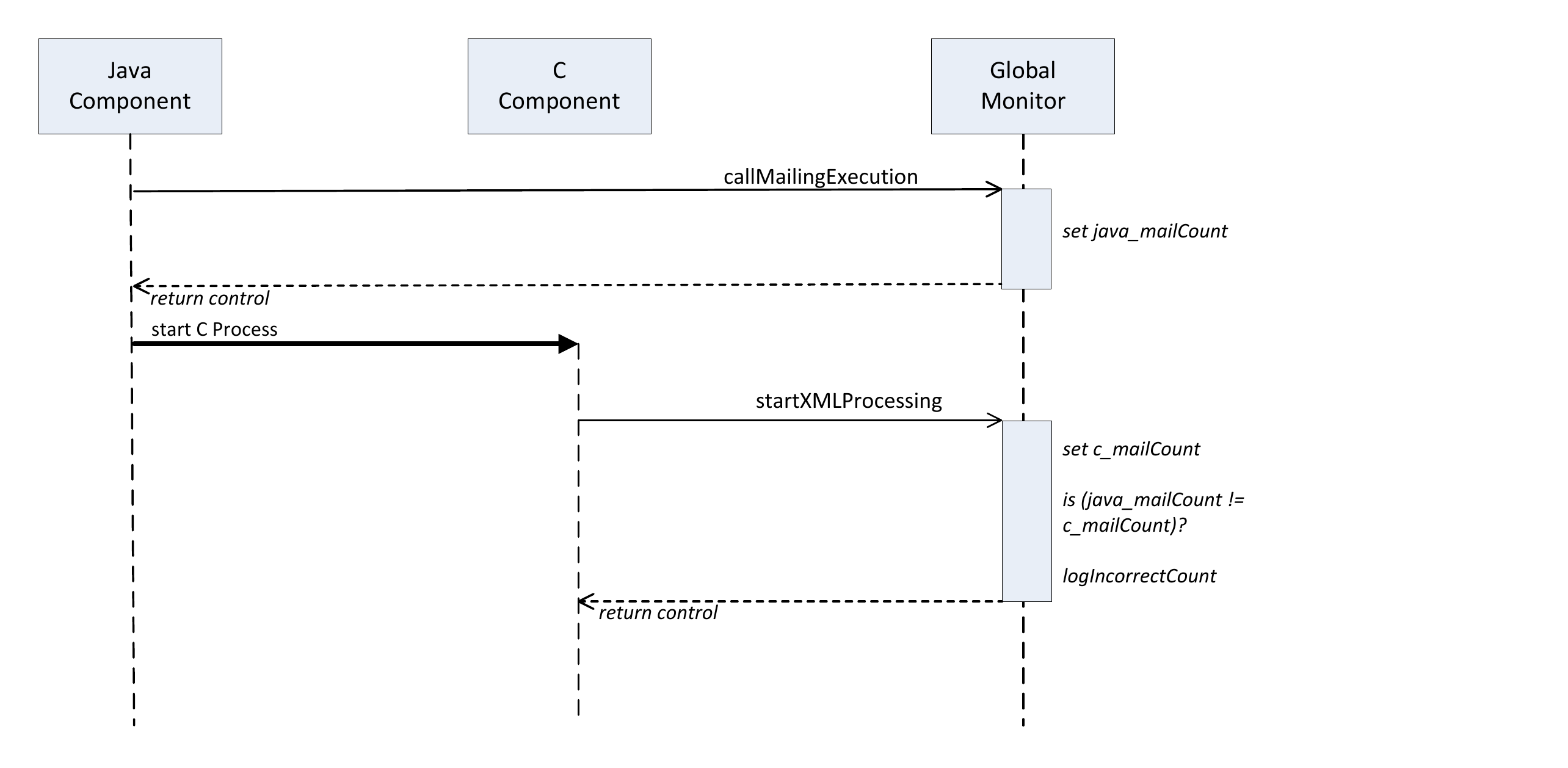}}
	\caption{Monitoring email shot subscribers}
	\label{fig:csequenceproperty1}
\end{figure}

A basic property that can be monitored to ensure the XML file has not been tampered is to ensure that the total number of recipients in the mailing list is the same within the C component as it is in the Java component. This may be expressed in terms of the following rule:
$$\begin{array}{l}
\mboxit{\texttt{c\_sendMails()}}  \mid  
  \mboxit{\texttt{(java\_mailCount != c\_mailCount)}} \mapsto 
  \mboxit{\texttt{logIncorrectCount}}; \\ 
\end{array}$$
where \texttt{c\_sendMails} is a system event occurring on the \OpenEmm C process at the point when mails are about to be sent; the rule condition specifies that the total number of mails being sent by the C process \texttt{c\_mailCount} must be equal to the total count of mails that was specified at the Java component. \texttt{logIncorrectCount} is the action taken by the monitor if the values are not equal. 
Fig.~\ref{fig:csequenceproperty1} shows the flow of control amongst the components involved in monitoring the specified property: the monitoring listeners instrumented within the Java and C components, and the global monitoring component. When the details of a mail shot are available, the Java component notifies the global monitor with the total number of subscribers. The value is stored as the monitor state \texttt{java\_mailCount}. 
Subsequently, when the C component receives the XML file, the total number of mail recipients is communicated to the global monitor which in turn compares it to \texttt{java\_mailCount}. Program~\ref{prg:monitoringrecipcount} shows the \newlarva specification required to generate this setup. In \ding{179} and \ding{180} the specification distinguishes between the events' component-sources while the main rule is specified in \ding{181}.

\begin{program}
\begin{newcode}
upon (newMailShot(mailshotID))  \{
    state \{ 
      monitorSide \{int java\_mailCount;  \} 
    \}
    events \{
        \ding{179} event@javaComponent callMailingExecution(mailshotID, javaSubsCount) = 
                                  \{MailShot.startExecution(mailshotID, javaSubsCount);\}
        \ding{180} event@cComponent startXMLProcessing(mailshotID, c\_mailCount) =
                                  \{parse\_receivers(mailshotID, c\_mailCount);\}
    \}
    conditions \{
      monitorSide \{ invalidMailCount(c\_mailCount) = java\_mailCount != c\_mailCount; \}
    \}    
    actions \{
      monitorSide \{ setJavaMailCount(javaSubsCount) = 
                             java\_mailCount == javaSubsCount; \}
      monitorSide \{ logIncorrectCount = ...\}
    \}
    rules \{
      callMailingExecution(mailshotID, javaSubsCount) \verb+\+ true \verb+->+ 
                                              setJavaMailCount(javaSubsCount);          
      \ding{181} startXMLProcessing(mailshotID, c\_mailCount) \verb+\+ invalidMailCount \verb+->+ 
                                              logIncorrectCount;                                               
    \}
  \}
\end{newcode}
\caption{Monitoring count of mailshot recipients}
\label{prg:monitoringrecipcount}
\end{program}

\begin{figure}[t]
	\centering
		\scalebox{.75}{
		    \includegraphics{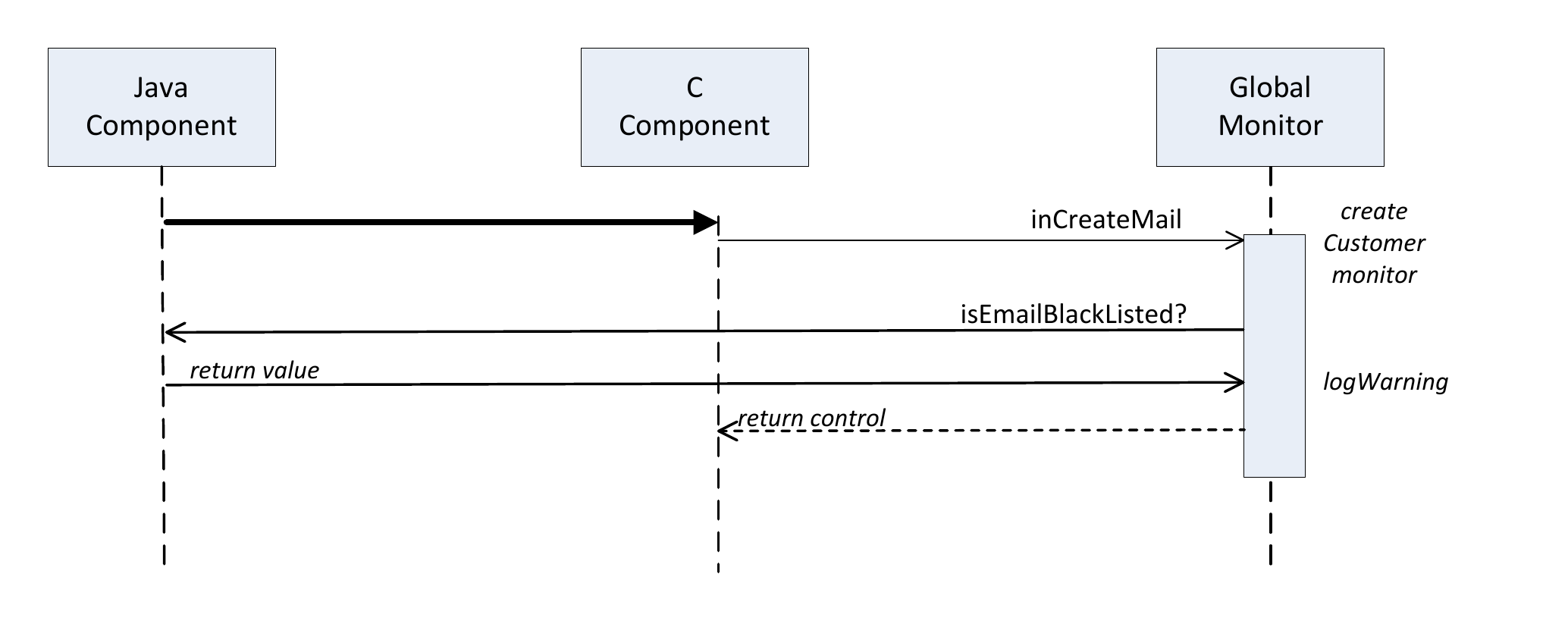}}
	\caption{Monitoring blacklisted recipients}
	\label{fig:csequenceproperty2}
\end{figure}

Another property which we monitored on \OpenEmm is that of ensuring that blacklisted users  are never sent an email. 
\OpenEmm adheres to this property by carrying out a filtering exercise on the mailing list recipients, leaving out any blacklisted recipients. 
However, if an email recipient is blacklisted while the mailing generation process is already running, there could be circumstances where the recipient is still included in the mailing list. 
Such an issue could be detected by setting up a \newlarva monitor which verifies that each recipient is still non-blacklisted at the time of being sent an email. 
Fig.\ \ref{fig:csequenceproperty2} depicts how the global runtime monitor can be notified upon the creation of a personalised email, \texttt{inCreateMail}, triggering the monitoring process to query the blacklist on the Java component, \texttt{isEmailBlackListed?}. Program \ref{prg:blacklisted} shows how the property of verifying users to be non-blacklisted can be specified in \newlarva. In particular note that the condition is specified as \texttt{systemSide@javaComponent} meaning that the condition is to be executed within the Java component and the result is then communicated to the global monitor. 


\begin{program}
\begin{newcode}
upon (newCustomer(custID))  \{
    events \{        
        event@cComponent inCreateMail(c\_custID) = \{create\_mail(c\_custID);\}
    \}
    conditions \{
      systemSide@javaComponent\{ isEmailBlacklisted(c\_custID) = 
                                   database query to check for c\_custID in blacklist ... \}
    \}    
    actions \{
      monitorSide \{ logBlacklisted = ...\}
    \}
    rules \{
      checkEmail = inCreateMail(mailshotID, c\_custID) \verb+\+ isEmailBlacklisted \verb+->+ 
                                              logEmailBlacklisted;          
    \}
  \}
\end{newcode}
\caption{Monitoring blacklisted recipients}
\label{prg:blacklisted}
\end{program}


The two properties described above have been successfully compiled by \newlarva and applied to \OpenEmm. 
Each specification script was processed by three compilers --- 
$(i)$ the standard \newlarva\ compiler that creates the global monitoring component, 
$(ii)$ the \newlarva\ language compiler in conjunction with the Java plugin, to create a component listener that was woven into the Java \OpenEmm code, and 
$(iii)$ the \newlarva\ language compiler in conjunction with the C plugin, to create a component listener that was woven into the C \OpenEmm code. \OpenEmm was installed on an Ubuntu operating system while the global monitor component was executed on a separate Windows machine. 


No performance tests were carried out during this case study, since the aim of our work was to study the interaction between different components running different technologies and the global runtime monitoring component. 
The possible performance improvements that can be achieved using the \newlarva framework are however discussed in our other work \cite{conf/sefm/ColomboFMP12}.


\section{Conclusions}

While a significant number of runtime verification frameworks have been proposed in the literature
\cite{Colombo:2009:DER:1614485.1614501,FJN+11bip,Meredith:2012:OMR:2215862.2215866,d'Amorim:2005:ERV:1083246.1083249,Kim:2004:JRA:972560.972567,Giannakopoulou:2001:RAL:891184}, these tools are normally restricted to support one particular programming language or technology and the effort required to support new languages is prohibitive. 
An exception is the MOP framework~\cite{Meredith:2012:OMR:2215862.2215866} whose architecture makes provision for the addition of new language plugins that can generate a MOP runtime monitor for a particular programming language. However, MOP does not support the monitoring of a single property across a system with multiple technologies and does not have an inbuilt concept of components. On the other hand, a runtime verification approach has been proposed for the BIP component framework \cite{FJN+11bip} which tackles issues specific to component-based systems. However, the work is positioned at a higher level of abstraction, focusing on the theoretical guarantees that are required to ensure sound and correct monitoring within BIP --- \ie the issue of multiple technologies has not been considered in this work.



The non-monolithic nature of component-based systems means that verification techniques have to be adapted to be applicable. In this paper, we have presented an extension of an existing tool, \newlarva \cite{conf/sefm/ColomboFMP12}, to handle the runtime verification of component-based systems. In particular, we have emphasised the need for the support of multiple-technologies used in such systems, with the resulting tool being easily extensible so as to handle new technologies. Although we have shown its applicability by deploying it on a third-party open-source system, we are currently looking into its use in an industrial setting. Furthermore, we are looking into ways of combining our runtime verification approach to the unit testing of component-based systems.

\bibliographystyle{eptcs}
\bibliography{refs}

\end{document}